# Analysis of Gaussian width of nucleons in semi-classical model.

Supriya Goyal and Rajeev K. Puri*
*Department of Physics, Panjab University, Chandigarh-160014, INDIA*
* email: rkpuri@pu.ac.in

## Introduction

Heavy-ion collisions at intermediate energies lead to final states containing many complex fragments. As the beam energy increases from few tens to several hundreds of MeV/nucleon, the multiplicity of final fragments increases steadily. The study of nuclear collisions at intermediate energies is primarily motivated by the unique possibilities for probing the physical properties of hot and dense nuclear matter. Among the various non-equilibrium effects [1], which play an important role in a realistic treatment of heavy-ion collisions, momentum dependent interactions (MDI) show most pronounced effect. Therefore, momentum dependent potentials have been implemented into the early dynamics as well as into the time dependent meson field approaches. The significant influence of MDI on the collective flow, particle production, rapidity distribution, anisotropy ratio, density, temperature, and stability of the nuclei etc. has been reported in the literature [1,2,3]. One of the common problems of the simulation of a heavy-ion reaction is the proper description of the ground state nuclei. The problem is more severe once the MDIs are included as it leads to an additional repulsion between the nucleons when boosted. The nucleus build up in the initial stage should be stable in its ground state as well as on a time scale comparable with the time span needed for the nucleus-nucleus collision. A lot of improvements are done in the literature on this context [4]. We plan to investigate the stability of the nuclei propagating with and without MDI in the framework of Quantum Molecular Dynamics (QMD) model [2,3]. Clusterization algorithm based on the spatial approach and commonly used i.e. Minimum Spanning Tree (MST) algorithm has been employed for the present study [2,3].

## Model

In the QMD model [2,4], each nucleon is represented by a coherent state of the form

$$\psi_i(\mathbf{r},\mathbf{p}_i(t),\mathbf{r}_i(t))=\frac{1}{(2\pi L)^{3/4}}\exp\left[\frac{i}{\hbar}\mathbf{p}_i(t)\mathbf{r}-\frac{(\mathbf{r}-\mathbf{r}_i(t))^2}{4L}\right]. \quad (1)$$

Here L is the Gaussian width with standard value of 1.08 fm$^2$. It signifies the root mean square radius of nucleons. The nucleons propagate according to the following classical equations of motion.

$$\frac{d\vec{p}_i}{dt}=-\frac{dH}{d\vec{r}_i} \quad and \quad \frac{d\vec{r}_i}{dt}=\frac{dH}{d\vec{p}_i}, \quad (2)$$

where the Hamiltonian is given by:

$$H=\sum_i\frac{p_i^2}{2m_i}+\sum_i V_i^{Loc}+V_i^{Coul}+V_i^{Yuk}+V_i^{MDI}. \quad (3)$$

Here $V_i^{Skyrme}$, $V_i^{Coul}$, $V_i^{Yuk}$, and $V_i^{MDI}$ are, respectively, the Skyrme, Coulomb, Yukawa and momentum dependent potential. The MDI is obtained by parameterizing the momentum dependence of the real part of the optical potential. The final form of the MDI potential reads as:

$$V^{MDI}=t_4\ln^2(t_5(\vec{p}_1-\vec{p}_2)^2+1)\delta(\vec{r}_1-\vec{r}_2). \quad (4)$$

Here $t_4$ = 1.57 MeV and $t_5$ = 5×10$^{-4}$ MeV$^{-2}$. The imaginary part of the potential is parameterized in terms of nucleon-nucleon cross section.

## Results and discussion

For the present analysis, the selected nuclei i.e. $^{12}$C, $^{40}$Ca, $^{93}$Nb, and $^{197}$Au are simulated with static soft equation of state (EoS) and the momentum dependent soft EoS (labelled as Soft and SMD, respectively). From Ref. [3], we noted that on including MDI, the main contribution of the emitted units goes as free nucleons, therefore for the present study; we have treated all the emitted units at the final step of 200fm/c as free nucleons. The standard value of the Gaussian wave packet width i.e. $L^{stand}$ = 1.08 fm$^2$ was taken in Ref. [3]. The impact of doubling the width of Gaussian wave packets (i.e. $L^{broad}$ = 2.16 fm$^2$) on the stability of the nuclei is shown in the fig. 1. In fig. 1, we display the time evolution of the mass of the largest fragment formed ($A^{max}$, calculated using MST algorithm) for the all the four cold nuclei mentioned earlier. Various lines are explained in the caption of the figure. We clearly see from the figure that $A^{max}$ decreases with time, but the decrease is more for SMD EoS compared to Soft EoS for $L^{stand}$ case. On doubling the width i.e. with $L^{broad}$, the size of $A^{max}$ matches with the true size of nuclei, both for Soft and SMD EoS. It means that nuclei having double width do not emit free nucleons for a long period of time necessary to study heavy-ion collisions. Also, the ground state properties of all the nuclei are also described well. In the low mass region, the obtained nuclei are less bound but stable. Heavy mass nuclei have proper binding energy and are stable (results are not shown here) [5].

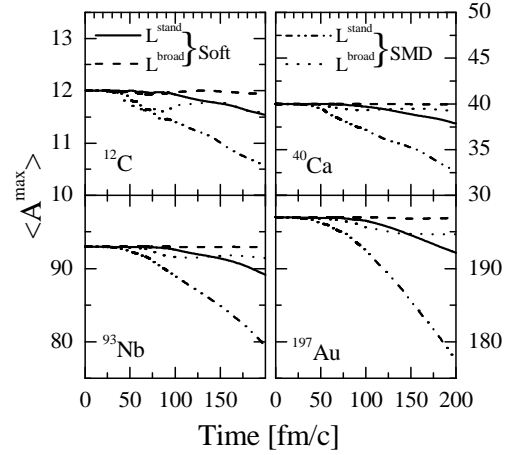

**Fig. 1** The time evolution of $A^{max}$ emitted from single cold nuclei of $^{12}$C, $^{40}$Ca, $^{93}$Nb, and $^{197}$Au. Results of Soft (SMD) with $L^{stand}$ and $L^{broad}$ are represented by solid (dash double dotted) and dashed (dotted) lines, respectively.

## Acknowledgments

This work is supported by grant from Department of Science and Technology (DST), Govt. of India.

## References:

[1] J. Aichelin, A. Rosenhauer, G. Peilert, H. Stocker, and W. Greiner, Phys. Rev. Lett. **58,** 1926 (1987); A. D. Sood and R. K. Puri, Phys. Rev. C **79,** 064618 (2009); S. Kumar, S. Kumar, and R. K. Puri, Phys. Rev. C **78,** 064602 (2008); S. Kumar, S. Kumar, and R. K. Puri, Phys. Rev. C **81,** 014611 (2010); S. Kumar, S. Kumar, and R. K. Puri, Phys. Rev. C **81,** 014601 (2010); Y. K. Vermani, J. K. Dhawan, S. Goyal, R. K. Puri, and J. Aichelin, J. Phys. G **37,** 015105 (2010).
[2] Ch. Hartnack *et al.*, Eur. Phys. J. A **1,** 151 (1998).
[3] Y. K. Vermani, S. Goyal, and R. K. Puri, Phys. Rev. C **79,** 064613 (2009).
[4] K. Niita *et al.*, Phys. Rev. C **52,** 2620 (1995); K. Abdel-Waged, Phys. Rev. C **71,** 044607 (2005); L. De Paula *et al.*, Phys. Lett. B **258,** 251 (1991).
[5] S. Goyal and R. K. Puri, Phys. Rev. C (2010) - to be submitted.